\documentstyle[preprint,aps]{revtex}
\begin{document}
\title{Eigenstates of linear combinations of  two  boson  creation 
and annihilation operators : An algebraic approach}
\author{P. Shanta, S. Chaturvedi and V.Srinivasan}
\address{School of Physics, University of Hyderabad\\
Hyderabad - 500 046, {\bf INDIA}}
\maketitle
\begin{abstract}
Eigenstates of the linear  combinations  $a^2+\beta  a^{\dagger2}$ 
and $ab+\beta a^\dagger  b^\dagger$  of  two  boson  creation  and 
annihilation operators  are  presented.  The  algebraic  procedure 
given here is based on the work of Shanta et al. [Phys. Rev. Lett. 
{\bf 72}, 1447, 1994] for constructing eigenstates of  generalized 
annihilation operators. Expressions for the overlaps of 
these states with the number states, the coherent states and the squeezed 
states are given in a closed form.
\end{abstract}
\vskip0.5in
\newpage
\noindent{\bf 1.~~ Introduction}

In a recent work$^1$ it was shown that for an operator $F$ made up 
of powers of a single bosonic annihilation operator or of products 
of commuting bosonic annihilation operators one  can  construct  a 
canonical conjugate $G^\dagger_i$ such that $[F,G^\dagger_i]=1$ on 
all states in a specific sector  $S_i$  of  the  Fock  space.  The 
eigenstates of the  generalized  annihilation  operators  $F$  and 
$G_i$ were constructed and it was shown that these coherent states 
come in pairs and are in some sense duals of each other. Thus, for 
instance, it was shown  that  the  cat  states$^2$  and  the  Yuen 
states$^3$  form  a  dual  pair  and  so  do  the  pair   coherent 
states$^{4-6}$   and   the   Caves-Shumaker    states.$^7$    This 
construction  has  also  been  recently   extended   to   deformed 
algebras.$^8$

In  this  work  we  show  that  a  two  fold  application  of  the 
construction developed in ref. 1 can also be  put  to  use  for  a 
purely algebraic construction of the eigenstates of  operators  of 
the type $F+\beta  F^\dagger$.  In  particular,  we  consider  two 
operators of this type.
\def\cf{{\cal F}}
\def\cg{{\cal G}}
\def\ad{a^\dagger}
\def\bd{b^\dagger}
\begin{itemize}
\item[(i)] $\cf_1=(a^2+\beta a^{\dagger2})~;~[a,\ad] = 1\,\,\,$
\item[(ii)] 
$\cf_2=(ab+\beta\ad\bd)~;~[a,\ad]=1~,~[b,\bd]=1~,~[a,b]=0\,\,.$
\end{itemize}

It   is   well   known   that   the    operators    ${1\over2}a^2, 
{1\over2}a^{\dagger2}$  and  ${1\over2}(\ad   a+{1\over2})$   when 
identified respectively with $K_-$,  $K_+$  and  $K_z$  furnish  a 
realization of  the  $su(1,1)$  algebra  $[K_z,K_\pm]=\pm  K_\pm$, 
$[K_-,K_+]=2K_z$. Similarly the  operators  $K_-=ab$,  $K_+=\ad\bd$ 
and  $K_z={1\over2}(a\ad+b\bd)$   provide   a   two   mode 
realization  of  the  same  algebra.  In  view  of  this,  one  is 
considering  here  the  eigenvalue  problem   for   the   operator 
$K_-+\beta K_+$. The eigenfunctions of the operator  $\cf_1$  have 
recently been constructed by  Nieto and Truax$^9$ and by Satya  Prakash  
and  Agarwal$^{10}$  by 
solving the appropriate  differential  euqation. The latter authors  
have  also 
investigated the non-classical aspects of these states. A similar 
analysis has been carried out for the  eigenfunctions  of  $\cf_2$ 
corresponding to the zero eigenvalue of  $a\ad-b\bd$.$^{11}$  Here 
we present a complete solution to the problem  using  an  entirely 
algebraic approach.
\vskip0.5cm
\noindent{\bf 2.~~ Eigenstates of The Operator $\cf_1$}

The solution of the eigenvalue problem
\begin{equation}
\cf_1\mid \psi> = \lambda\mid \psi>~~~~~ \cf_1=(a^2+\beta  a^{\dagger  2}) 
\,\,\,\,,
\end{equation}
involves two steps
\begin{enumerate}
\item Construction of the state $\mid \psi>_o$ annihilated by $\cf_1$
\begin{equation}
\cf_1\mid \psi>_o = 0\,\,\,\,.
\end{equation}
\item Construction of the canonical  conjugate  $\cg$  to  $\cf_1$ 
satisfying $[\cf_1, \cg^\dagger]=1$.
\end{enumerate}

\noindent{\it Construction of the state $\mid \psi>_o$ annihilated  by 
$\cf_1$}:

We rewrite (2) as
\begin{equation}
a^2\mid \psi>_o = -\beta a^{\dagger2}\mid \psi>_o\,\,\,\,,
\end{equation}
and apply $a^2$ on both  sides.  The  resulting  equation  can  be 
written as
\begin{equation}
F\mid \psi>_o = -\beta\mid \psi>_o~~~,~~F={1\over(n_1+1)(n_a+2)} a^4 ~~, ~~ 
n_a\equiv \ad a\,\,\,\,.
\end{equation}

Thus the task of solving (3) reduces to constructing the solutions 
of the eigenvalue problem (4). It must be borne in mind  that  the 
solutions  of  (3)  satisfy  (4)  but  not   vice   versa.   After 
constructing the solutions of (4) one has to discard  those  which 
are not solutions of (3).

To find the eigenstates of $F$ in (4) we follow the  procedure  of 
ref. 1. The states annihilated by $F$ are  $\mid 0>$,  $\mid 1>$,  $\mid 2>$ 
and $\mid 3>$. Successive applications of $F^\dagger$  on  these  four 
generate the four sectors
\begin{equation}
S_0 = \{\mid 4n>\}, ~ S_1=\{\mid 4n+1>\},~ S_2=\{\mid 4n+2>\}, ~ S_3=\{\mid 4n+3>\} 
\,\,\,\,,
\end{equation}
where $n=0,1,\ldots\ldots$. As shown in Appendix A, the canonical conjugates 
 $G^\dagger_i$ of $F$, satisfying $[F, G^\dagger_i] = 1$ in the sector $S_i$ 
 are 

\begin{eqnarray}
G_o^\dagger & = & {a^{\dagger4}\over4} {1\over(n_a+3)}~~~ ,\\
G_1^\dagger & = & {a^{\dagger4}\over4} {1\over(n_a+4)}~~~ ,\\
G_2^\dagger  &  =  &   {a^{\dagger4}\over4}   {(n_a+2)\over(n_a+4) 
(n_a+3)}~~~ ,\\
G_3^\dagger  &  =  &   {a^{\dagger4}\over4}   {(n_a+1)\over(n_a+4) 
(n_a+3)}~~~ .
\end{eqnarray}

The general solution of (4) may thus be written as
\begin{eqnarray}
\mid \psi>_o & = & C_o\exp(-\beta  G\dagger_o)\mid 0>  +  C_1  \exp(-\beta 
G^\dagger_1)\mid 1> + C_2\exp(-\beta G^\dagger_2)\mid 2>\nonumber\\
& + & C_3\exp(-\beta G^\dagger_3)\mid 3>\,\,\,\,.
\end{eqnarray}
Of these four independent solutions of (4) the  latter  two,  which 
are specific linear combinations of  the  states  in  the  sectors 
$S_2$ and $S_3$ respectively, are  not  solutions  of  (3)  as  can 
readily be verified. The general solution of (2) is thus
\begin{equation}
\mid \psi>_o  =   C_o\exp(-\beta  G^\dagger_o)\mid 0>   +   C_1\exp(-\beta 
G^\dagger_1)\mid 1>\,\,\,.
\end{equation}
This  completes  the  first  step  towards  the  solution  of  the 
eigenvalue problem (1).

\noindent{\it   Construction   of    the    canonical    conjugate 
$\cg^\dagger$ of $\cf_1$}

We begin by noticing that the canonical conjugates  $\sl g^\dagger_i$, 
$i=0,1$ of $a^2$
\begin{eqnarray}
{\sl g}^\dagger_o & = & {a^{\dagger2}\over2} {1\over(n_a+1)}\,\,\,\,,\\
{\sl g}^\dagger_1 & = & {a^{\dagger2}\over2} {1\over(n_a+2)}\,\,\,\,,
\end{eqnarray}
satifying
\begin{equation}
[a^2, {\sl g}^\dagger_i] = 1\,\,\,\,,
\end{equation}
in the sectors
\begin{equation}
S_o = \{\mid 2m>\}~~,~~ S_1=\{\mid 2m+1>\}~~,~~m = 0,1\ldots\ldots\,\,\,,
\end{equation}
respectively also satisfy
\begin{equation}
[a^{\dagger2}, {\sl g}^\dagger_i] = 4{\sl g}^{\dagger2}_i\,\,\,\,.
\end{equation}
This  suggests  the  following  form  for   $\cg^\dagger_i$,   the 
canonical conjugate of $\cf_1\equiv a^2+\beta a^{\dagger2}$, we  are 
looking for
\begin{equation}
\cg^\dagger_i = \sum_{n=1}^\infty b_n({\sl g}^\dagger_i)^n\,\,\,.
\end{equation}
On requiring that
\begin{equation}
[a^2+\beta a^{\dagger2}, \cg_i^\dagger] = 1\,\,\,,
\end{equation}
we find that
\begin{equation}
b_{2m} = 0 ~~~, ~~~ b_{2m+1} = {(-4\beta)^m\over(2m+1)} \,\,\,\,,
\end{equation}
so that
\begin{equation}
\cg^\dagger_i  =   {1\over\sqrt{4\beta}}   \tan^{-1}(\sqrt{4\beta} 
{\sl g}^\dagger_i)\,\,\,\,.
\end{equation}
The completes the construction  of  $\cg^\dagger_i$.  The  general 
solution of the eigenvalue problem (1) is given by
\begin{equation}
\mid \psi> = C_o\mid \psi,e> + C_1\mid \psi,0>\,\,\,\,\,,
\end{equation}
where
\begin{eqnarray}
\mid \psi,e>   &   =    &    \exp(\lambda    \cg^\dagger_o)\exp(-\beta 
G^\dagger_o)\mid 0 >\,\,\,\,,\\
\mid \psi,o>   &   =    &    \exp(\lambda    \cg^\dagger_1)\exp(-\beta 
G^\dagger_1)\mid 1>\,\,\,\,.
\end{eqnarray}

\noindent{\bf 3.~~ Overlap between the eigenstates of $\cf_1$  and 
the squeezed vacuum $\mid \mu>$:}

As noted in [1] the squeezed vacuua
\begin{equation}
\mid \mu,e>  =  \exp\mu  a^{\dagger2}\mid 0>  ~~;~~  \mid \mu,o>   =   \exp\mu 
a^{\dagger2}\mid 1>\,\,\,,
\end{equation}
in the even and odd sectors are respectively eigenstates of  ${\sl g}_o$ 
and ${\sl g}_1$.
\begin{equation}
{\sl g}_o\mid \mu,e> = \mu\mid \mu,e>~~~ ;~~~ {\sl g}_1\mid \mu,o> = \mu\mid \mu,o>\,\,\,\,,
\end{equation}
Calculation of the overlap between  the  state  $\mid \psi>$  and  the 
squeezed vacuua would become rather easy if one could express  the 
operators on the RHS of (22) and (23) entirely in terms  of  ${\sl 
g}^\dagger_i$. The operators ${\cal G}^\dagger_i$ are  related  to 
${\sl g}^\dagger_i$ through (20). Further, using the fact that
\begin{equation}
G^\dagger_o  =   {\sl   g}_o^{\dagger2}(n_a+1)   \qquad   ;   \qquad 
G_1^\dagger =  {\sl g}_1^{\dagger2}(n_a+2)\,\,\,\,\,,
\end{equation}
it can  be shown that
\begin{eqnarray}
\exp(-\beta    G^\dagger_o)\mid 0>     &     =     &     (1+4\beta{\sl 
g}_o^{\dagger2})^{-1/4}\mid  0>\,\,\,\,\,,\\
\exp(-\beta    G^\dagger_1)\mid 1>     &     =     &     (1+4\beta{\sl 
g}_1^{\dagger2})^{-3/4}\mid  1>\,\,\,\,\,.
\end{eqnarray}
The derivation of (27) entails expanding the exponential and  \\
(a) using $ [{\sl   g}_o^{\dagger2}(n_a+1)]^q = ({\sl   g}_o^{\dagger2})^q
{(n_a+1)(n_a+4)\cdots (n_a+4(q-3))}$ \\
(b) applying ${(n_a+1)(n_a+4)\cdots (n_a+4(q-3))}$ on $|0>$ and \\
(c) and summing up the resulting series using 
 $$  
 (1+x)^{-\alpha} = \sum_{n=0}^{\infty} (-1)^n \frac{\Gamma(\alpha +n)}
{\Gamma(\alpha)} \frac{x^n}{n!} \,\,\,.
$$
The derivation of (28) involves exactly the same steps.The states $\mid \psi,e>$ and $\mid \psi,o>$ can thus be written as
\begin{eqnarray}
\mid \psi,e>    &    =     &     \exp\left({\lambda\over\sqrt{4\beta}} 
\tan^{-1}(\sqrt{4\beta} {\sl g}^\dagger_o)\right)  (1+4\beta  {\sl 
g}^{\dagger2}_o)^{-1/4} \mid 0>\,\,\,\,,\\
\mid \psi,o>    &    =     &     \exp\left({\lambda\over\sqrt{4\beta}} 
\tan^{-1}(\sqrt{4\beta} {\sl g}^\dagger_1)\right)  (1+4\beta  {\sl 
g}^{\dagger2}_1)^{-3/4} \mid 1>\,\,\,\,,
\end{eqnarray}
from which it follows that
\begin{eqnarray}
<\mu,e\mid \psi,e>    &    =     &     \exp\left({\lambda\over\sqrt{4\beta}} 
\tan^{-1}(\sqrt{4\beta} \mu^*)\right)  (1+4\beta  \mu^{*2})^{-1/4} 
\,\,\,\,,\\
<\mu,o\mid \psi,o>    &    =     &     \exp\left({\lambda\over\sqrt{4\beta}} 
\tan^{-1}(\sqrt{4\beta} \mu^*)\right)  (1+4\beta  \mu^{*2})^{-3/4} 
\,\,\,\,.
\end{eqnarray}
These finite expressions for the ovelaps are valid for $\mid\mu\sqrt{\beta}\mid
< 1/2 $.

\noindent{\bf 4. Overlap between the eigenstates of  ${\cal  F}_1$ 
and the coherent states $\mid \alpha>$:}

The calculate the  overlap  between  $\mid \psi>$  and  the  coherent 
states $\mid \alpha>$, it proves convenient to express the operators on 
the RHS of (29) and (30) in terms  $a^\dagger$.  Simple  algebraic 
manipulations outlined in Appendix B, enable us  to  rewrite  (29) 
and (30) as follows 
\begin{eqnarray}
\mid \psi,e>     &      =      &      \exp\left(-{i\over2}\sqrt{\beta} 
a^{\dagger2}\right)               M\left({1\over4}               - 
{i\lambda\over4\sqrt{\beta}}   ,   {1\over2}    ,    i\sqrt{\beta} 
a^{\dagger2}\right) \mid 0>\,\,\,\,,\\
\mid \psi,o>     &      =      &      \exp\left(-{i\over2}\sqrt{\beta} 
a^{\dagger2}\right) M\left({3\over4}               - 
{i\lambda\over4\sqrt{\beta}}   ,   {3\over2}    ,    i\sqrt{\beta} 
a^{\dagger2}\right) \mid 1>\,\,\,\,.
\end{eqnarray}
With  $\mid \psi,e>$  and  $\mid \psi,o>$  written  in   this   way,   one 
immediately obtains
\begin{eqnarray}
<\alpha\mid \psi,e> & = &  \exp\left(-{i\over2}\sqrt{\beta}\alpha^{*2} 
\right)M\left({1\over4}      -      {i\lambda\over4\sqrt{\beta}}, 
{1\over2},   i\sqrt{\beta}\alpha^{*2}\right)   \exp(-\mid \alpha\mid ^2/2) 
\,\,\,\,,\\
<\alpha\mid \psi,o> & = &  \alpha^*\exp\left(-{i\over2}\sqrt{\beta}\alpha^{*2} 
\right)M\left({3\over4}      -      {i\lambda\over4\sqrt{\beta}}, 
{3\over2},   i\sqrt{\beta}\alpha^{*2}\right)   \exp(-\mid \alpha\mid ^2/2) 
\,\,\,\,,
\end{eqnarray}
where $M(a,b,z)$ denote the confluent hypergeometric  functions. 
From (35) and (36) one can readily calculate the $Q$-function  for 
these states.

\noindent{\bf 5. Overlap with number states:}

We first consider $\mid \psi,e>$. Expanding the RHS of (33) in powers of
$a^\dagger$ we obtain

\begin{equation}
\mid \psi,e> = {\Gamma({1\over2})\over\Gamma({1\over4}-{i\lambda\over4
\sqrt{\beta}})}
\sum_{k=0}^\infty\sum_{l=0}^\infty \frac{1}{k!}{\left(-{1\over2}
\right)}^k {\Gamma({1\over4}-{i\lambda\over4\sqrt{\beta}}+l) \over 
\Gamma(l+{1\over2})l!} (i{\sqrt\beta})^{k+l} (a^\dagger)^{2(k+l)}\mid 0>
\,\,\,,
\end{equation}
which, in turn, yields
\begin{equation}
<2n\mid \psi,e> = {\Gamma({1\over2})\over\Gamma({1\over4}-{i\lambda\over4
\sqrt{\beta}})}{\left(\frac{-i\sqrt{\beta}}{2}\right)}^n \sqrt{2n!}
\sum_{l=0}^n {\Gamma({1\over4}-{i\lambda\over4\sqrt{\beta}}+l) \over 
\Gamma(l+{1\over2})l!(n-l)!} (-2)^l \,\,\,.
\end{equation}
The expression on the RHS may be expressed in terms of the hypergeometric
functions as follows
\begin{equation}
<2n\mid \psi,e> = {\left(\frac{-i\sqrt{\beta}}{2}\right)}^n \frac{\sqrt{(2n)!}}
{n!} F\left(-n,{1\over4}-{i\lambda\over4\sqrt{\beta}};\frac{1}{2};2\right)
\,\,\,.
\end{equation}
Similarly
\begin{equation}
<2n+1\mid \psi,o> = {\left(\frac{-i\sqrt{\beta}}{2}\right)}^n 
\frac{\sqrt{(2n+1)!}}
{n!} F\left(-n,{3\over4}-{i\lambda\over4\sqrt{\beta}};\frac{3}{2};2\right)
\,\,\,.
\end{equation}

\noindent{\bf 6. Coordinate space wave function:}

Using (33) and (34) we can easily derived the expressions for  the 
coordinate space wave functions  for  the  states  $\mid \psi,e>$  and 
$\mid \psi,o>$. Details are given in Appendix C. The  (un  normalized) 
wave functions turn out to be 
\begin{eqnarray}
<x\mid \psi,e> & =  &  \exp\left[-{1\over2}\left({1+i\sqrt{\beta}\over 
1-i\sqrt{\beta}}\right)x^2\right]  M\left({1\over4}-{i\lambda\over 
4\sqrt{\beta}},    {1\over2}    ,     {2i\sqrt{\beta}\over1+\beta} 
x^2\right) \,\,\,\,,\\
<x\mid \psi,o> & =  & x \exp\left[-{1\over2}\left({1+i\sqrt{\beta}\over 
1-i\sqrt{\beta}}\right)x^2\right]  M\left({3\over4}-{i\lambda\over 
4\sqrt{\beta}},    {3\over2}    ,     {2i\sqrt{\beta}\over1+\beta} 
x^2\right) \,\,\,\,.
\end{eqnarray}

\noindent{\bf 7. Eigenstates of the operator ${\cal F}_2$:}

We next consider the eigenvalue problem for  the  operator  ${\cal 
F}_2$.
\begin{equation}
{\cal F}_2\mid \phi> = \lambda\mid \phi> \qquad ; \qquad {\cal F}_2 = (ab + 
\beta a^\dagger b^\dagger)\,\,\,\,.
\end{equation}
As before this task can be broken up into two steps - construction 
of the state $\mid \phi>_o$ annihilated by ${\cal F}_2$
\begin{equation}
{\cal F}_2\mid \phi>_o = 0\,\,\,\,,
\end{equation}
and the construction of the canonical conjugate ${\cal G}^\dagger$ 
of ${\cal F}_2$.

We rewrite (43) as
\begin{equation}
ab\mid \phi>_o = -\beta a^\dagger b^\dagger\mid \phi>_o\,\,\,\,,
\end{equation}
and apply $ab$ on  both  sides.  The  resulting  equation  can  be 
written as
\begin{equation}
F\mid \phi>_o    =    -\beta    \mid \phi>_o     \qquad,     \qquad     F= 
{1\over(n_a+1)(n_b+1)} a^2b^2\,\,\,\,.
\end{equation}

Now the states annihilated by $F$ are $\mid 0,p>$,  $\mid q,0>$,  $p\ge0$, 
$q>0$,  and  $\mid 1,q>$  and  $\mid q,1>;  p\ge   1,   q>1$.   Successive 
applications  of  $F^\dagger$  on  these  generate  the  following 
sectors
\begin{eqnarray*}
S_{o,p} & = & \{\mid 2n,2n+p>\}  ,  p\ge  0\qquad  ;  \qquad  S_{q,o}  = 
\{\mid 2n+q,2n>\}, q>0 \,\,;\\
S_{1,p} & = & \{\mid 2n,+1,2n+p>\}  ,  p\ge  1\qquad  ;  \qquad  S_{q,1}  = 
\{\mid 2n+q,2n+1>\}, q>1 \,\,,
\end{eqnarray*}
where $n=0,1,\cdots$. As shown in Appendix A, the canonical  conjugates  
$G^\dagger_i$  of $F$, satisfying
\begin{equation}
[F,G^\dagger_i] =1\,\,\,\,,
\end{equation}
in these sectors are found to be
\begin{eqnarray}
G^\dagger_o    &    = &       {1\over2}       a^{\dagger2}b^{\dagger2} 
{(n_a+2)\over(n_a+2)(n_b+2)}\,\,\,\,,\\
G^\dagger_1    &    = &       {1\over2}       a^{\dagger2}b^{\dagger2} 
{(n_b+2)\over(n_a+2)(n_b+2)}\,\,\,\,,\\
G^\dagger_2    &    = &       {1\over2}       a^{\dagger2}b^{\dagger2} 
{(n_a+1)\over(n_a+2)(n_b+2)}\,\,\,\,,\\
G^\dagger_3    &    = &       {1\over2}       a^{\dagger2}b^{\dagger2} 
{(n_b+1)\over(n_a+2)(n_b+2)}\,\,\,\,.
\end{eqnarray}

The general solution of (46) may then be written as
\begin{eqnarray}
\mid \phi>_o    &    =    &    \sum_{p=o}^\infty    C_{o,p}\exp(-\beta 
G^\dagger_o)\mid 0,p>   +   \sum_{q=1}^\infty   C_{1,q}    \exp(-\beta 
G^\dagger_1) \mid q,0>\nonumber\\
   &    =    &    \sum_{p=1}^\infty    C_{2,p}\exp(-\beta 
G^\dagger_2)\mid 1,p>   +   \sum_{q=2}^\infty   C_{3,q}    \exp(-\beta 
G^\dagger_3) \mid q,1>\,\,\,\,\,.
\end{eqnarray}
The last two terms in (52) satisfy (46) but not  (45)  and  should 
therefore be discarded. The general solution of (45) is thus
\begin{equation}
\mid \phi>_o = \sum_{p=o}^\infty C_{o,p} \exp(-\beta G_o^\dagger)\mid 0,p> 
+ \sum_{q=1}^\infty C_{1,q} \exp(-\beta G^\dagger_1)\mid q,0>\,\,\,.
\end{equation}
The  completes  the  first  step  towards  the  solution  of   the 
eigenvalue problem  (53).  Notice  that  $\mid \phi>_o$  is  a  linear 
combination of states in the sectors $S_{o,p}$ and  $S_{q,o}$.  To 
construct  ${\cal  G}^\dagger$,  we  notice  that  the   canonical 
conjugates of $ab$ in the  sectors  $S_{o,p}$  and  $S_{q,o}$  are 
respectively given by
\begin{eqnarray}
{\sl g}^\dagger_o & = & a^\dagger b^\dagger {1\over(n_b+1)}\,\,\,,
\\
{\sl g}^\dagger_1 & = & a^\dagger b^\dagger {1\over(n_a+1)}\,\,\,,
\end{eqnarray}
satisfying
\begin{equation}
[ab, {\sl g}_i^\dagger] = 1\,\,\,\,.
\end{equation}
These canonical conjugates also satisfy 
\begin{equation}
[a^\dagger b^\dagger, {\sl g}^\dagger_i]  =  {\sl  g}^{\dagger2}_i 
\,\,\,\,.
\end{equation}
As before, taking ${\cal G}^\dagger_i$ to be of the form
\begin{equation}
{\cal    G}^\dagger_i    =     \sum_{n=1}^\infty     b_n     ({\sl 
g}_i^\dagger)^n\,\,\,\,,
\end{equation}
and requiring that
\begin{equation}
[ab+\beta a^\dagger b^\dagger, {\cal G}^\dagger_i] = 1\,\,\,.
\end{equation}
we find that
\begin{equation}
b_{2m} = 0 \qquad,\qquad b_{2m+1} = {(-\beta)^m\over(2m+1)}\,\,\,,
\end{equation}
so that 
\begin{equation}
{\cal G}_i^\dagger =  {1\over\sqrt{\beta}}  \tan^{-1}(\sqrt{\beta} 
g_i^\dagger)\,\,\,.
\end{equation}
The general solution of the eigenvalue problem (43) is then  given 
by
\begin{equation}
\mid \phi> = \sum_{p=o}^\infty C_{o,p} \mid \phi;0,p> +  \sum_{q=1}^\infty 
C_{1,q} \mid \phi;q,0>\,\,\,,
\end{equation}
where
\begin{eqnarray}
\mid \phi;0,p>  &\equiv&  \exp(\lambda{\cal   G}^\dagger_o)\exp(-\beta 
G^\dagger_o)\mid 0,p>\,\,\,\,,\\
\mid \phi;q,0>  &\equiv&  \exp(\lambda{\cal   G}^\dagger_1)\exp(-\beta 
G^\dagger_1)\mid q,0>\,\,\,\,.
\end{eqnarray}

\noindent{\bf 8. Overlap between the eigenstates of  ${\cal  F}_2$ 
and the generalised Caves Schumaker States:}

It was noted in ref 1. that the (un normalized) states
\begin{eqnarray}
\mid \mu;0,p> & = & \exp \mu  a^\dagger  b^\dagger\mid 0,p>\,\,\,;\,\,\,\, 
p=0,1,\cdots ,\\
\mid \mu;q,0> & = & \exp \mu  a^\dagger  b^\dagger\mid q,0>\,\,\,;\,\,\,\, 
q=1,2\cdots .
\end{eqnarray}
(hereafter referred to as the generalised Caves-Schumaker  states) 
are eigenstates of ${\sl  g}^\dagger_o$  and  ${\sl  g}^\dagger_1$ 
respectively. These  states,  which  contain  the  Caves-Schumaker 
state $\exp\mu a^\dagger b^\dagger\mid 0,0>$, are the counterparts  of 
the states $\mid \mu,e>$ and $\mid \mu;o>$ of  section  3  and  have  been 
studied in detail in ref 12. To express the overlap of these  with 
the states $\mid \phi;0,p>$ and $\mid \phi;q,0>$, we expresses the  latter 
in terms ${\sl g}_o^\dagger$ and $g^\dagger_1$ as follows
\begin{eqnarray}
\mid \phi;0,p> & = & \exp\left(  {\lambda\over\sqrt{\beta}}  \tan^{-1} 
(\sqrt{\beta}    {\sl    g}_o^{\dagger})\right)    (1+\beta    
{\sl g}_o^{\dagger2})^{-(p+1)/2} \mid 0,p>\,\,\,\,,\\
\mid \phi;q,0> & = & \exp\left(  {\lambda\over\sqrt{\beta}}  \tan^{-1} 
(\sqrt{\beta}    {\sl    g}_1^\dagger)\right)    (1+\beta    {\sl 
g}_1^{\dagger2})^{-(q+1)/2} \mid q,0>\,\,\,\,.
\end{eqnarray}
The steps involved in deriving (67) and (68) are exactly the same as those used
in obtaining (27) and (28).
From the expressions one readily obtains
\begin{eqnarray}
<\mu;0,p|\phi;0,p>  &  =  &   \exp\left({\lambda\over\sqrt{\beta}} 
\tan^{-1}(\sqrt{\beta}\mu^*)\right)(1+\beta\mu^{*2})^{-(p+1)/2}\, 
\,\,,\\
<\mu;q,0|\phi;q,0>  &  =  &   \exp\left({\lambda\over\sqrt{\beta}} 
\tan^{-1}(\sqrt{\beta}\mu^*)\right)(1+\beta\mu^{*2})^{-(q+1)/2}\, 
\,\,.
\end{eqnarray}
These finite expressions for the ovelaps are valid for $|\mu\sqrt{\beta}|< 1$

\noindent{\bf 9.~~ Overlap between the eigenstates of ${\cal F}_2$ 
and the coherent states $|\gamma,\delta>$: the $Q$ functions}

Using the results given in Appendix B, one finds that $|\phi,0,p>$ 
and $|\phi;q,0>$ can  be  expressed  in  terms  of  the  operators 
$a^\dagger$ and $b^\dagger$ as follows
\begin{eqnarray}
|\phi;0,p>  &  =  &   \exp(-i\sqrt{\beta}a^\dagger 
b^\dagger) M\left({p+1\over2} - {i\lambda\over2\sqrt{\beta}} \,,\, 
p+1\,,\,2i\sqrt{\beta} a^\dagger b^\dagger\right) |0,p>\,\,\,,\\
|\phi;q,0>  &  =  &   \exp(-i\sqrt{\beta}a^\dagger 
b^\dagger) M\left({q+1\over2} - {i\lambda\over2\sqrt{\beta}} \,,\, 
q+1\,,\,2i\sqrt{\beta} a^\dagger b^\dagger\right) |q,0>\,\,\,.
\end{eqnarray}
The overlaps of these states with the  two  mode  coherent  states 
$|\gamma,\delta>$ are therefore given by
\begin{eqnarray}
<\gamma,\delta|\phi;0,p>     &     =     &     \exp(-i\sqrt{\beta} 
\gamma^*\delta^*)\exp(-(|\gamma|^2+|\delta|^2)/2)\nonumber\\
&\times & M\left({p+1\over2} - {i\lambda\over2\sqrt{\beta}}  \,\,, 
\,\, p+1\,\,,\,\, 2i\sqrt{\beta} \gamma^*\delta^*\right)\,\,\,,\\
<\gamma,\delta|\phi;q,0>     &     =     &     \exp(-i\sqrt{\beta} 
\gamma^*\delta^*)\exp(-(|\gamma|^2+|\delta|^2)/2)\nonumber\\
&\times & M\left({q+1\over2} - {i\lambda\over2\sqrt{\beta}}  \,\,, 
\,\, q+1\,\,,\,\, 2i\sqrt{\beta} \gamma^*\delta^*\right)\,\,\,.
\end{eqnarray}
From these expressions one can readily calculate the corresponding 
$Q$-functions.

\noindent{\bf 10. Overlap with number states:}

We first consider $\mid \phi;0,p>$. Expanding the RHS of (71) in powers of
$a^\dagger b^\dagger$ we obtain

\begin{equation}
\mid \phi;0,p> = {\Gamma(p+1)\over\Gamma({p+1\over2}-{i\lambda\over2
\sqrt{\beta}})}
\sum_{k=0}^\infty\sum_{l=0}^\infty \frac{1}{k!}(-1)^k (2)^l {\Gamma({p+1\over2}
-{i\lambda\over2\sqrt{\beta}}+l) \over 
\Gamma(l+p+1)l!} (i{\sqrt\beta})^{k+l} (a^\dagger b^\dagger)^{(k+l)}\mid 0,p>
\,\,\,,
\end{equation}
which, in turn, yields
\begin{equation}
<n,n+p\mid \phi;0,p> = {\Gamma(p+1)\over\Gamma({p+1\over2}-{i\lambda\over2
\sqrt{\beta}})}(-i\sqrt{\beta})^n \sqrt{n!(n+p)!}
\sum_{l=0}^n {\Gamma({p+1\over2}-{i\lambda\over2\sqrt{\beta}}+l) \over 
\Gamma(l+p+1)l!(n-l)!} (-2)^l \,\,\,.
\end{equation}
The expression on the RHS may be expressed in terms of the hypergeometric
functions as follows
\begin{equation}
<n,n+p\mid \phi;0,p> = (-i\sqrt{\beta})^n \frac{\sqrt{n!(n+p)!}}
{n!} F\left(-n,{p+1\over2}-{i\lambda\over2\sqrt{\beta}};p+1;2\right)
\,\,\,.
\end{equation}
Similarly
\begin{equation}
<n+q,n\mid \phi;q,0> = (-i\sqrt{\beta})^n \frac{\sqrt{n!(n+q)!}}
{n!} F\left(-n,{q+1\over2}-{i\lambda\over2\sqrt{\beta}};q+1;2\right)
\,\,\,.
\end{equation}

\noindent{\bf 11. Concluding Remarks}

To  conclude,  we  have  given  a  purely  algebraic  method   for 
constructing   the   eigenstates   of   the    operators    ${\cal 
F}_1=(a^2+\beta a^{\dagger2})$ and ${\cal F}_2=(ab+\beta a^\dagger 
b^\dagger)$. The operator ${\cal F}_2$  on  making  the  canonical 
transformation
\begin{equation}
a=(c+id)/\sqrt{2}\,\,,\,\,    b=(c-id)/\sqrt{2}\,\,;\,\,    a^\dagger    = 
(c^\dagger+id^\dagger)/\sqrt{2}\,\,,\,\,b^\dagger                = 
(c^\dagger-id^\dagger)/\sqrt{2}\,\,\,,
\end{equation}
can be written as
\begin{equation}
{\cal   F}_2    =    {1\over2}    (c^2+d^2)    +    \beta{1\over2} 
(c^{\dagger2}+d^{\dagger2})\,\,\,\,,
\end{equation}
and is therefore the sum of two operators of  ${\cal  F}_1$  type. 
Thus in constructing the eigenstates of ${\cal F}_2$ given by (43) 
one has also constructed the eigenstates of  the  operator  ${\cal 
F}_2$ given by (80).  The  construction  presented  here  directly 
expresses the eigenstates of these operators  in  the  exponential 
form i.e., as states obtained by applying exponentials of  certain 
operators on the appropriate ``vacuua''. As  is  well  known,  the 
Yuen states and the  Caves-Schumaker  states  are  eigenstates  of 
linear combinations of creation and  annihilation  operators.  The 
states   constructed   here   may   be   considered   as   natural 
generalisations of these in the sense that they are eigenstates of 
operators which involve linear combinations of squares or products 
creation and annihilation operators and, like  the  Yuen  and  the 
Caves-Shumaker states, may find  useful  applications  in  quantum 
optics.
\newpage
\noindent{\bf Appendix - A}

The canonical conjugates $G^\dagger_i$ of any single mode annihilation 
operator of the form $f(n_a) a^p$ where $f(x)$ has no zeros at integer values of
$x$ (including zero) are given by the formula$^1$
$$
G^\dagger_i =\frac{1}{p} F^\dagger \frac{1}{FF^\dagger} (n_a+p-i)\,\,\,; \,\,
i=0,\cdots,p-1 \,\,\,. 
\eqno(A.1)
$$
Thus,for instance, for the operator $F$ in (4) given by
$$
F = \frac{1}{(n_a+1)(n_a+2)} a^4 \,\,\,\,,  
\eqno(A.2)
$$
one has 
$$
FF^\dagger = \frac{(n_a+4)(n_a+3)}{(n_a+1)(n_a+2)} \,\,\,\,,  \eqno(A.3)
$$
and hence
$$
G^\dagger_i = \frac{1}{4} a^{\dagger 4} \frac{(n_a+4-i)}{(n_a+4)(n_a+3)} \,\,\,.
\eqno(A.4)
$$
Setting $ i = 0,1,2,3 $ one obtains the expressions in (6)-(9). Similarly, for 
$F=a^2$, one obtains (12) and (13).

 Consider now a two mode annihilation operator consisting of products 
of single mode annihilation operators of the above type.
$$
F=F_1(a) F_2(b) \,\,\,,
\eqno(A.5)
$$
where
$$
F_1(a) = f_1(n_a) a^k \,\,\,;\,\,\, F_2(b) = f_2(n_b) b^l\,\,\,.
\eqno(A.6)
$$
The vacuua of $F$ are $|i,p>,\,\,; i= 0,\cdots,k-1$ and $|q,j>,\,\,; j=0,\cdots
,l-1$. The canonical conjugates of $F$ in the sectors built on $|i,p>$ are given 
by 
$$
G^\dagger_i =\left[\frac{1}{k} F_1^\dagger \frac{1}{F_1F_1^\dagger} (n_a+k-i)
\right] \left[F_2^\dagger \frac{1}{F_2F_2^\dagger}\right] \,\,\,.
\eqno(A.7)
$$
Similarly, the canonical conjugates of $F$ in the sectors built on 
$|q,j>\,\,, j= 0, \cdots,l-1$ are given by
$$
G^\dagger_j =\left[\frac{1}{l} F_2^\dagger \frac{1}{F_2F_2^\dagger} (n_b+l-j)
\right]\left[F_1^\dagger \frac{1}{F_1F_1^\dagger}\right] \,\,\,.
\eqno(A.8)
$$
Thus, for instance, for the operator F in (46) one has
$$
F_1 = \frac{1}{n_a+1}a^2 \,\,\,;\,\,\,F_2 = \frac{1}{n_b+1}b^2 \,\,\,,
\eqno(A.9)
$$
for which
$$
F_1F_1^\dagger = \frac{(n_a+2)}{(n_a+1)} \,\,\,;\,\,\, F_2F_2^\dagger =
 \frac{(n_b+2)}{(n_b+1)} \,\,\,,
\eqno(A.10)
$$
and hence, for the sectors built on $|i,p>$
$$
G_i^\dagger = \frac{1}{2}a^{\dagger2}\frac{1}{(n_a+2)}(n_a+2-i) b^{\dagger2}
\frac{1}{(n_b+2)} \,\,\,.
\eqno(A.11)
$$
Setting $i=0,1$, one obtains (48) and (49). Similarly (A.8) yields (50) and (51). The same considerations as above, applied to the operator $F=ab$, yield (50)
and (55). 
 
\noindent{\bf Appendix - B}

In  this  Appendix  we  show  that  the  expressions  for   states 
$|\psi;e>$ and $|\psi,o>$ the given by (29) and (30) which involve 
the exponential of an inverse $\tan$ function respectively can  be 
rewritten as in (33) and (34). Similarly, we show that the  states 
$|\phi;0,p>$ and $|\phi;q,0>$ in (67) and (68) can be rewritten  as 
in (71) and (72).

\noindent Consider, for instance, $|\phi;0,p>$
$$ 
|\phi;0,p>   =   \exp\left({\lambda\over\sqrt{\beta}}    \tan^{-1} 
(\sqrt{\beta} {\sl g}^\dagger_o)\right){(1+\beta {\sl g}^{\dagger2}_o})^{-
(p+1)/2} |0,p>\,\,\,\,. \eqno(B.1)
$$ 
Defining
$$ 
z\equiv \lambda/\sqrt{\beta} \qquad  \mbox{and}  \qquad  x  \equiv 
\sqrt{\beta} {\sl g}^\dagger_0\,\,\,\,, \eqno(B.2)
$$ 
(B.1) can be written as 
$$ 
|\phi;0,p>   =    \exp(z    \tan^{-1}    x)    (1+x^2)^{-(p+1)/2} 
|0,p>\,\,\,. \eqno(B.3)
$$ 
Using the identity
$$
\tan^{-1} x = {1\over2i} \ln \left({1+ix\over1-ix}\right)\,\,\,\,,
$$
one finds that
$$ 
\exp(z \tan^{-1}x)(1+x^2)^{-(p+1)/2} = (1+ix)^{-(p+1)}  \left(1  - 
{2ix\over1+ix}\right)^{(iz-(p+1))/2}\,\,\,. \eqno(B.4)
$$ 
Using 
$$ 
(1+{\em x})^{-\alpha} =  \sum_{n=0}^\infty  {\Gamma(\alpha+n)\over 
\Gamma(\alpha)} {(-{\em x})^n\over n!}\,\,\,\,,\eqno(B.5)
$$ 
the RHS of (B.4) can be expanded  in  powers  of  $x$  as   
\begin{eqnarray*} 
\lefteqn{\exp(z \tan^{-1}x)(1+x^2)^{-(p+1)/2} = }\\
& & \sum_{q=0}^\infty \sum_{k=0}^\infty 
{(2i)^q(-i)^k\Gamma(q+(p+1-iz)/2) \Gamma(p+q+k+1)\over 
\Gamma((p+1-iz)/2) \Gamma(p+q+1) q! k!} x^{k+q}\,\,\,. 
&({\it B}.6)\cr
\end{eqnarray*} 
The action of $x^{k+q}$ on $|0,p>$ yields
\begin{eqnarray*}  
x^{k+q}|0,q> & = &(\sqrt{\beta}a^\dagger)^{k+q}{\left(b^\dagger\frac{1}{n_b+1}\right)}^{k+q}|0,p> \cr
&=&  {\Gamma(p+1)\over\Gamma(k+q+p+1)}   (\sqrt{\beta} 
a^\dagger b^\dagger)^{k+q} |0,p>\,\,\,\,. 
&({\it B}.7)\cr
\end{eqnarray*} 
Using (B.6) and (B.7) in (B.1) one obtains
\begin{eqnarray*}
|\phi;0,p>  & = &  \left[\sum_{k=0}^\infty   {(-i\sqrt{\beta}a^\dagger 
b^\dagger)^k\over k!}\right]  \cr
& \times & \left[{\Gamma(p+1)\over\Gamma((p+1-iz)/2)} 
\sum_{q=0}^\infty   {\Gamma(q+(p+1-iz)/2)\over    \Gamma(p+q+1)q!} 
(2i\sqrt{\beta} a^\dagger b^\dagger)^q\right]|0,q>\,\,, &({\it B}.8)\cr
& = & \exp(-i\sqrt{\beta} a^\dagger b^\dagger) M\left({p+1\over2} 
- {i\lambda\over2\sqrt{\beta}}\,\,,\,\,p+1\,\,,\,\,2i\sqrt{\beta} 
a^\dagger b^\dagger\right) |0,p>\,\,, 
&({\it B}.9)\cr
\end{eqnarray*}
which is the same as (67).\\
Following the same procedure one can derive (33), (34) and (72) from (29), 
(30) and (68).

\noindent{\bf Appendix -  C}

In this Appendix we derive the expressions in (41) and (42) for the coordinate 
space wave functions for $|\psi,e>$ and $|\psi,o>$. We first consider $|\psi;e>$. 
From (33) we have
\begin{eqnarray*}
&& <x|\psi,e>\cr
&=&<x|\exp\left(-{i\over2}\sqrt{\beta}a^{\dagger2}\right)
M\left({1\over4}-{i\lambda\over4\sqrt{\beta}}\,\,,\,\,{1\over2}\,\,, 
\,\,i\sqrt{\beta}a^{\dagger2}\right)|0>\,\,\,,\cr
& =& {\Gamma({1\over2})\over\Gamma({1\over4}-{i\lambda\over4\sqrt{\beta}})}
\sum_{l=0}^\infty {\Gamma({1\over4}-{i\lambda\over4\sqrt{\beta}}+l) \over 
\Gamma(l+{1\over2})l!} (-1)^l<x|(-i\sqrt{\beta} a^{\dagger2})^l 
\exp\left(-{1\over2}\sqrt{\beta} a^{\dagger2}\right)|0>\,\,\,,\cr
& =& {\Gamma({1\over2})\over\Gamma({1\over4}-{i\lambda\over4\sqrt{\beta}})}
\sum_{l=0}^\infty {\Gamma({1\over4}-{i\lambda\over4\sqrt{\beta}}+l) \over 
\Gamma(l+{1\over2})l!} (-1)^l \left[{\partial^l\over\partial\mu^l}
<x|\exp(-i\sqrt{\beta}\mu a^{\dagger2})|0><\right]_{\mu=1/2}\,. 
&({\it C}.1)\cr
\end{eqnarray*}
To proceed further, we need to know $<x|\exp(-i\sqrt{\beta}\mu a^{\dagger2})|0>$.
This can be done by \\
(a) expanding the exponential \\
(b) using the fact that
$$
<x|a^{\dagger2m} |0> = \frac{1}{[\Gamma({\frac{1}{2}})]^{\frac{1}{2}}} 
\exp\left(-{x^2\over2}\right) (-2)^m m! L_m ^{-\frac{1}{2}}(x^2) \,\,\,,
\eqno(C.2)
$$
(c) and recognizing the series thus obtained as the generating function of the associated Laguerre functions. This leads to 
$$
<x|\exp(-i\sqrt{\beta} \mu a^{\dagger2})|0> = {1\over\sqrt{\Gamma({1\over2}) 
(1-2i\mu\sqrt{\beta})}} \exp\left(-{x^2\over2}\right) 
\exp\left({2i\mu\sqrt{\beta}x^2\over2i\mu\sqrt{\beta}-1}\right)\,\,.\eqno(C.3)
$$
We rewrite this expression as
\begin{eqnarray*} 
<x|\exp(-i\sqrt{\beta} \mu a^{\dagger2})|0> & = & 
{1\over\sqrt{\Gamma({1\over2})(1-2i\mu\sqrt{\beta})}} 
\exp\left(-{1\over2} {(1+i\sqrt{\beta})\over(1-i\sqrt{\beta})} x^2\right) \\
& \times & \exp\left({i\sqrt{\beta}(1-2\mu)x^2\over (i\sqrt{\beta}-1) 
(2i\mu\sqrt{\beta}-1)}\right)\,,
&({\it C}.4)\cr
\end{eqnarray*}
and define
$$
z = \left(\frac{2i\sqrt{\beta}}{1-i\sqrt{\beta}}\right)\left(\mu -\frac{1}{2}
\right) \,\,\,,
\eqno(C.5)
$$
to obtain
\begin{eqnarray*}
&&\left[{\partial^l\over\partial\mu^l} <x|\exp(-i\sqrt{\beta}  \mu 
a^{\dagger2})|0>\right]_{\mu=1/2}\cr
&=&{1\over\sqrt{\Gamma({1\over2})(1-i\sqrt{\beta})}}   \exp 
\left(-{1\over2}{(1+i\sqrt{\beta})\over(1-i\sqrt{\beta})}      x^2 
\right)   \left({2i\sqrt{\beta}\over1-i\sqrt{\beta}}\right)^l
\left[{\partial^l\over\partial z^l}     {1\over\sqrt{(1-z)}}\right.\cr
&\times &       \left.  \exp\left(\frac{z}{z-1}\left(\frac{x^2}{1-i\sqrt{\beta}}\right)\right)\right]_{z=0}\,,\cr
&  =  &   {1\over\sqrt{\Gamma({1\over2})(1-i\sqrt{\beta})}}   \exp 
\left(-{1\over2}{(1+i\sqrt{\beta})\over(1-i\sqrt{\beta})}      x^2 
\right)   \left({2i\sqrt{\beta}\over1-i\sqrt{\beta}}\right)^l   l! 
{L_l}^{-\frac{1}{2}}\left({x^2\over1-i\sqrt{\beta}}\right)\,\,\,. \cr
&&    
&({\it C}.6)\cr
\end{eqnarray*}
Substituting  this  in (C.1) we get 
\begin{eqnarray*}
<x|\psi,e> &=& {\sqrt{\Gamma({1\over2})}
\over\Gamma({1\over4}-{i\lambda\over4\sqrt{\beta}})} 
{1\over\sqrt{(1-i\sqrt{\beta})}} \exp\left(-{1\over2} 
{(1+i\sqrt{\beta})\over(1-i\sqrt{\beta})} x^2\right)\cr
& \times & \sum_{l=0}^\infty {\Gamma({1\over4} - 
{i\lambda\over4\sqrt{\beta}}+l)\over\Gamma(l+{1\over2})} 
\left({-2i\sqrt{\beta}\over1-i\sqrt{\beta}}\right)^l 
{L_l}^{-\frac{1}{2}}\left({x^2\over1-i\sqrt{\beta}}\right)\,\,\,.
&({\it C}.7)\cr
\end{eqnarray*}
Putting the expansion
$$
{L_l}^{-\frac{1}{2}}\left({x^2\over1-i\sqrt{\beta}}\right) = \sum_{m=0}^l 
{\Gamma(l+{1\over2})\over\Gamma(m+{1\over2})\Gamma(l-m+1)m!} 
\left({-x^2\over1-i\sqrt{\beta}}\right)^m\,\,\,,\eqno(C.8)
$$
for the Laguerre polynomials in (C.8) and rearranging  the  double 
series we get
\begin{eqnarray*}
<x|\psi,e> &=& {\sqrt{\Gamma({1\over2})}
\over\Gamma({1\over4}-{i\lambda\over4\sqrt{\beta}})} 
{1\over\sqrt{(1-i\sqrt{\beta})}} \exp\left(-{1\over2} 
{(1+i\sqrt{\beta})\over(1-i\sqrt{\beta})} x^2\right)\cr
& \times & \sum_{m=0}^\infty {1\over\Gamma(
(m+{1\over2})m!} 
\left({2i\sqrt{\beta}x^2\over(1-i\sqrt{\beta})^2}\right)^m \cr
&\times & \sum_{k=0}^\infty 
{\Gamma({1\over4}-{i\lambda\over4\sqrt{\beta}}+m+k)\over k!}
\left({-2i\sqrt{\beta}\over1-i\sqrt{\beta}}\right)^k\,\,\,. 
&({\it C}.9)\cr
\end{eqnarray*}
The sum over $k$ is easily carried out using (B.5) and gives
\begin{eqnarray*}
&& \sum_{k=0}^\infty {\Gamma({1\over4}-{i\lambda\over4\sqrt{\beta}} + 
m  +   k)\over   k!}   \left({-2i\sqrt{\beta}\over1-i\sqrt{\beta}} 
\right)^k \cr
&& \qquad \qquad\qquad\qquad = \Gamma({1\over4}-{i\lambda\over4\sqrt{\beta}} + m)
\left({1-i\sqrt{\beta}\over1+i\sqrt{\beta}}\right)^{({1\over4}   - 
{i\lambda\over4\sqrt{\beta}}+m)} \,.
&({\it C}.10)\cr
\end{eqnarray*}
Substituting this in (C.9) and recognizing the infinite series  as 
that for a confluent hypergeometric function, we finally obtain
$$
<x|\psi,e>                                                       = 
\exp\left(-{1\over2}{(1+i\sqrt{\beta})\over(1-i\sqrt{\beta})} 
x^2\right) M\left({1\over4} -\frac{i\lambda}{4\sqrt{\beta}}\,,\,{1\over2}\,,\,
{2i\sqrt{\beta}x^2\over1+\beta}\right)\,\,,\eqno(C.11)
$$
where we have omitted factors independent of $x$. Proceedings in exactly the
same way one can derive the  expression (42) for $<x\mid \psi;o>$.
\newpage
\noindent{\bf References}
\begin{enumerate}
\item P. Shanta, S. Chaturvedi, V. Srinivasan,  G.S.  Agarwal  and 
      C.L. Mehta, Phys. Rev. Lett. {\bf72}, 1447 (1994).
\item M. Hillery, Phys. Rev. A{\bf36}, 3796 (1987); C.C. Gerry and 
      E.E. Hach III, Phys. Lett. A{\bf174}, 185 (1993).
\item H.P. Yuen, Phys. Rev. A{\bf13}, 2226 (1976).
\item B. Bhaumik, K. Bhaumik and  B.  Dutta-Roy,  J.  Phys.  Math. 
      A{\bf9}, 1507 (1976).
\item A.O. Barut and C. Girardello, Commun. Math.  Phys.  {\bf21}, 
      41 (1971).
\item G.S. Agarwal, J. Opt. Soc. Am. B{\bf5}, 1940 (1988).
\item C.M. Caves and B.L. Schumaker, Phys.  Rev.  A{\bf31},  3068, 
      3093 (1985); K. W\'odkiewicz and J.H. Eberly, J.  Opt.  Soc. 
      Am. B{\bf2}, 458 (1985).
\item P. Shanta, S. Chaturvedi, V. Srinivasan and R.  Jagannathan, 
      J. Phys. A{\bf27}, 1 (1994).
\item M.M. Nieto and R. Truax , Phys. Rev. Lett. {\bf71}, 2843 (1993).
\item G. Satya Prakash and G.S. Agarwal, Phys. Rev. A{\bf50},  4258 
      (1994).
\item G. Satya Prakash and G.S. Agarwal, Phys. Rev. A {\bf52}, 2335 (1995).
\item C.C. Gerry, J. Opt. Soc. am. B{\bf8}, 685 (1991);  L.  Giles 
      and P.L. Knight J. Mod. Opt. {\bf 39}, 1411 (1992).
\end{enumerate}
\end{document}